\documentclass[preprint,aps,showpacs]{revtex4}

\usepackage{graphicx}
\usepackage{dcolumn}
\usepackage{bm}

\begin{document}

\title{Dispersion Management of Ultraslow Light in a Bose-Einstein Condensate via Trap Curvature}
\author{Devrim Tarhan$^1$, Se\c{c}kin Sefi$^2$, and \"O. E. M\"ustecapl{\i}o\u{g}lu$^{3,4}$
} \affiliation{$^1$Harran University, Department of Physics, 63300
Osmanbey Yerle\c{s}kesi, \c{S}anl\i{}urfa, Turkey}
\affiliation{$^2$Department of Physics, Istanbul Technical
University, Maslak 34469, Istanbul, Turkey} \affiliation{$^3$
Ko\c{c} University, Department of Physics, Rumelifeneri Yolu,
34450 Sar{\i}yer, Istanbul, Turkey} \affiliation{$^4$Institute of
Quantum Electronics, ETH Zurich Wolfgang-Pauli-Strasse 16, CH-8093
Zurich, Switzerland}

\date{\today}
\email{dtarhan@harran.edu.tr}
\begin{abstract}
One dimensional propagation of ultraslow optical pulses in an
atomic Bose-Einstein condensate taking into account the dispersion
and the spatial inhomogeneity is investigated. Analytical and
semi-analytical solutions of the dispersive inhomogeneous wave
equation modeling the ultraslow pulse propagation are developed
and compared against the standard wave equation solvers based upon
Cranck-Nicholson and pseudo-spectral methods. The role of
curvature of the trapping potential of the condensate on the
amount of dispersion of the ultraslow pulse is pointed out.
\end{abstract}

\pacs{03.75.Nt, 42.50.Gy, 81.05.Ni} \maketitle
%
%
%
%

\newpage
\section{Introduction}

Possibility of dramatic reduction of speed of light, as
demonstrated in an atomic Bose-Einstein condensate (BEC)
\cite{slowlight-exp1} using electromagnetically induced
transparency (EIT) \cite{eit1}, opens the way towards practical
realization of quantum optical information storage
\cite{qmemory1,qmemory2,qmemory3,qmemory4,liu}. In BECs, ultraslow
group velocities allow for storage times of coherent optical
information of the order of few microseconds. To enhance the
storage capacity and duration it is necessary to inject multiple
pulses to be simultaneously present in the condensate. Major
limiting factor against this goal is the narrow EIT window and
group velocity dispersion of the slow pulses.

Propagation of the ultraslow wave packet in one dimensional
inhomogeneous dispersive atomic condensate can be described in
terms of the slowly varying pulse envelope \cite{tarhan1,tarhan2}
\begin{equation} \label{eq:pulse}
\frac{\partial E}{\partial z} + \alpha(z) E + \frac{1}{v_g(z)}
\frac{\partial E}{\partial t} + i \,b_{2}(z) \frac{\partial^2
E}{\partial t^2}  = 0,
\end{equation}
where $\alpha(z)$ is the pulse attenuation factor; $v_g(z)$ is the
group velocity, and $b_{2}(z)$ is the group velocity dispersion.
The third order dispersion is found to be much smaller and
neglected \cite{tarhan2}. Previous studies are based upon either
analytical solutions obtained for an effective homogeneous region
at the center of the condensate where the dispersion is highest or
numerical propagation of the pulse with the standard methods such
as Cranck-Nicolson or pseudo-spectral schemes \cite{tarhan2}.

Present work aims to solve Eq. (\ref{eq:pulse}) analytically so
that optimum experimental conditions can be efficiently determined
to reduce the dispersive effects to enhance coherent information
storage capacity of the condensate. In addition to exact
analytical expressions of the pulse envelopes, an approximate
method of determination of the pulse envelope by polynomial or
Gaussian function fitting to condensate density profile is
introduced. The density profile only contributes as an integrand
to the optical parameters so that the fitting can be done with
very high accuracy for a simple and fast evaluation of the
required integrals. Besides, the spatial dependence of the
coefficients in the wave function is due to the inhomogeneous
condensate profile for which non-condensate, thermal component
makes a small contribution at low temperatures. The practical
semi-analytical approach outlined as above is tested against
Cranck-Nicolson and pseudo-spectral wave equation solvers and
shown to be highly efficient. The approach is employed to
investigate the effect of the curvature of the trapping potential
of the condensate, translated to the curvature of the dielectric
function of the condensate, on the amount of the dispersion of the
optical pulse.

The paper is organized as follows: In Sec. II, optical properties
of an atomic BEC system under EIT conditions is briefly reviewed.
In Sec. III, analytical and semi-analytical solutions are
developed for the wave equation. Numerical schemes used against
the semi-analytical solution are described in Sec. IV. Results are
discussed in Sec.V. Finally, we conclude in Section VI.
\section{Optical properties of an atomic BEC under EIT conditions}
\label{sec:opticalresponse}

Atomic condensates can be considered as two component objects,
composed of a condensate cloud in a thermal background so that one
can write its density being
\begin{eqnarray}
\rho(\vec{r})=\rho_c(\vec{r})+\rho_{T}(\vec{r}),
\end{eqnarray}
where \cite{naraschewski}
\begin{eqnarray}
\rho_c(\vec{r}) &=& \frac{\mu-V(r)}{U_{0}} \Theta(\mu-V(r)) \Theta(T_c-T),\\
\rho_{T}(\vec{r}) &=& \frac{g_{3/2} (z e^{-\beta V})}{
\lambda_T^3}.
\end{eqnarray}
Here $U_0=4\pi\hbar^2a_{s}/m$, $m$ is atomic mass, and $a_s$ is
the atomic s-wave scattering length. $\Theta(.)$ is the Heaviside
step function, $g_{n}(x)=\Sigma_{j} x^j/j^n$, $\lambda_T$ is the
thermal de Br\"{o}glie wavelength, $\beta=1/k_{B}T$, and $T_C$ is
the critical temperature. We assume an external trapping potential
in the form $V(\vec{r})=(1/2) m (\omega_r^2 r^2+\omega_z^2 z^2)$
with $\omega_r$ the radial trap frequency and $\omega_z$ the axial
trap frequency in the z direction. $\mu$ is the chemical
potential. At temperatures below $T_c$, the chemical potential
$\mu$ is determined by $\mu(T)=\mu_{TF}(N_0/N)^{2/5}$, where
$\mu_{TF}$ is the chemical potential evaluated under Thomas-Fermi
approximation. The condensate fraction is evaluated by
\cite{naraschewski} $N_0/N=1-x^3-s \zeta(2) /
\zeta(3)x^2(1-x^3)^{2/5}$, with $x=T/T_c$, and $\zeta$ is the
Riemann-Zeta function. The scaling parameter $s$ is given by
$s=\mu_{TF}/k_BT_C$.

EIT susceptibility \cite{eit2} for an atomic BEC of atomic density
$\rho$ can be expressed as
$\chi(\vec{r},\omega)=\rho(\vec{r})\chi_1(\omega)$ with
\begin {equation}
\label{chieit} \chi_1(\omega) = \frac{|\gamma|^2}{\epsilon_0
\hbar} \frac{{\rm i}(\Gamma_2/2-{\rm i} \Delta )}{[(\Gamma_2/2
-{\rm i}\Delta)(\Gamma_3/2 -{\rm i} \Delta) + \Omega_c^2/4]},
\end {equation}
where $\Delta=\omega-\omega_0$ is the detuning of the probe field
frequency $\omega$ from the atomic resonance $\omega_0$.
$\Omega_c$ is the Rabi frequency of the control field; $\gamma$ is
the dipole matrix element for the probe transition. $\Gamma_2$ and
$\Gamma_3$ denote the dephasing rates of the atomic coherence.

Through the atomic density, the optical response of the atomic
condensate becomes spatially inhomogeneous, so does the parameters
in the wave equation (\ref{eq:pulse}). They are determined by
\cite{harris}
\begin{eqnarray}
\alpha(\vec{r}) &=& -\frac{i \pi}{\lambda} \rho(\vec{r})\chi_1(\omega_{0})\nonumber,\\
\frac{1}{v_g(\vec{r})} &=& \frac{1}{c} -
\frac{\pi}{\lambda}\rho(\vec{r})\left. \frac{\mathrm{d}
\chi_1}{\mathrm{d}\omega}\right|_{\omega_{0}},\label{eq:vg}\\
b_{2}(\vec{r}) &=& \frac{\pi}{2\lambda}\rho(\vec{r})
\left[\left.\frac{\mathrm{d}^{2}\chi_1}{\mathrm{d}\omega^2}\right|_{\omega_{0}}\right].\nonumber
\end{eqnarray}
Thus the spatial dependence of all the optical parameters comes
solely from the axial density profile of the condensate. In the
following section we shall exploit the fact that it is only the
density profile that determines the local optical properties of
the condensate to develop an exact analytical solution of the wave
Eq.(\ref{eq:pulse}).
\section{Analytical and Semi-Analytical Methods for ultraslow pulse propagation}
\label{sec:analytical}

For a uniform density medium the Eq.\ref{eq:pulse} can be solved
analytically \cite{bahaa}. The complex envelope of the incident
wave is Gaussian pulse $E(0,t)=e^{-t^2/\tau_0^2}$. The initial
pulse, after propagating in the medium of length $L$, is then
found to be delayed with respect to a reference pulse propagating
in vacuum by $t_d=L/v_g$. The final width of the pulse is
$\tau(L)=\tau_0\sqrt{1+(L/z_0)^2}$ with $\tau_0$ is the initial
temporal width of the pulse, and $z_0=-\pi\tau_0^2/b_2$. For $L\gg
z_0$ we get $\tau(L)=|b_2|L/\pi\tau_0$. Experimentally measured
group velocity is defined by $v_g=L/t_d$, where the effective
axial length of the medium is evaluated by
$L=\left[(4\pi/N)\int_0^\infty\,r\mathrm{d}r\int_0^\infty\,\mathrm{d}z
z^2\rho(r,z)\right]^{1/2}$.

For a non-uniform medium, the wave equation (\ref{eq:pulse}) with
spatial dependent coefficients can be solved by using some basic
differential equation solving methods. In the typical slow light
experiments a small pin hole is introduced to couple incoming
light with the condensate guide and axial propagation is enforced.
In the subsequent discussions paraxial effects shall be ignored
and strictly one dimensional propagation will be considered by
taking $\vec{r}=(x,y,z)=(0,0,z)$. Fourier transforming Eq.
(\ref{eq:pulse}) from $t$ space to $w$ space gives
\begin{equation}
\label{diff} \frac{\mathrm{d} {\cal E}}{\mathrm{d} z} + \alpha(z)
{\cal E} - \frac{ {\rm i}w}{v_g(z)} {\cal E} -w^2{\rm i} b_{2}(z)
{\cal E} = 0
\end{equation}
where ${\cal E}={\cal
E}(z,w)\equiv(1/\sqrt{2\pi})\int_{-\infty}^{\infty}
E(z,t)\exp{({\rm i}wt)}\mathrm{d}\,t$. Solution for the equation
in $w$ space becomes
\begin{equation}
\label{integral1} {\cal E}(z,w)={\cal E}_0\;\exp{\left[\int^z
\left({\rm i}w^2 b_{2}(z') + {\rm i}w /v_g(z') - \alpha(z')
\right) \mathrm{d}\,z'\right]},
\end{equation}
where ${\cal E}_0=(1/\sqrt{2\pi})\int_{-\infty}^{\infty}
\exp{(-(t-t_0)^2/2\tau_0^2)}\exp{({\rm i}wt)}\mathrm{d}t=\tau_0
\exp{(-w^2\tau_0^2/2)}$. Transforming back to $t$ space we find
\begin{equation}
 E(z,t)=\frac{\tau_0}{\sqrt{2 f_1(z)}} \exp{\left[-\int^z
\alpha(z')\mathrm{d}z'-\frac{f_2^2(z)}{4f_1(z)}\right]},
\end{equation}
where
\begin{eqnarray}
f_1(z)&=&\left(\frac{\tau_0^2}{2}-\int^z b_{2}(z')\mathrm{d}z'\right),\\
f_2(z)&=&\left(t-t_0-\int^z \frac{1}{v_g(z')}\mathrm{d}z'\right).
\end{eqnarray}
Writing the trap potential with a variable curvature parameter
$\kappa$ such that $V=\kappa z^2$, density profile of 1D
Bose-Einstein condensate becomes $\rho(z) = [(\mu-V(z))/U_{0}]
\Theta(\mu-\kappa z^2) + g_{3/2} (\exp{(-\beta(\kappa z^2-\mu))})/
\lambda_T^3.$ To calculate the pulse envelope analytically, it is
necessary to be able to evaluate the single integral
\begin{equation}
 N(z)= \int_{-\infty}^z \rho(z^{'})\mathrm{d}z^{'}=N_0(z)+N_T(z),
 \label{totaldens}
\end{equation}
with $N_0(z)=\int_{-\sqrt{\mu / \kappa}}^{z}
\rho_0(z^{'})\mathrm{d}z^{'}$ and $N_T(z)=\int_{-\infty}^z
\rho_T(z^{'})\mathrm{d}z^{'}$.
$N_0$ makes the dominant contribution in the condensate region.
Let us define an auxiliary function for notational simplicity such
that
\begin{eqnarray}
F(z)=\frac{ 1}{ \lambda_T^3} \sum_{j=1}^{\infty}\sqrt{\frac{\pi }{
\beta \kappa}}\frac{
 e^{\beta \mu j}}{j^2} \phi(\sqrt{2 \beta \kappa j}z),
\end{eqnarray}
where $\phi(x)=1/(\sqrt{\pi}) \int_{-\infty}^x
e^{-u^2/2}\mathrm{d}u$ is the normal cumulative distribution
function which can be expressed in terms a tabulated special
function, error function as
$\phi(x)=1/2(1+\mathrm{erf}(x/\sqrt{2}))$. It is now possible to
express the results of the integral as follows
\begin{eqnarray}
N(z) = \cases {F(z) & $z<-|z_0|$ \cr
\frac{2}{3U_0}\sqrt{\frac{\mu^3}{\kappa}} + \frac{\mu z - \kappa
z^3/3}{U_0} + F(-|z_0|) & $z\in (-|z_0|,|z_0|)$ \cr
\frac{4}{3U_0}\sqrt{\frac{\mu^3}{\kappa}} + F(z) & $z>|z_0|$}.
\end{eqnarray}
Here $|z_0|=\sqrt{\mu/\kappa}$. Finally we can rewrite optical
pulse parameters in terms of $N(z)$ so that
\begin{eqnarray}
f_1(z)&=&\left(\frac{\tau_0^2}{2}- \frac{\pi}{2 \lambda}
\left.\frac{\mathrm{d}^{2}\chi_1}{\mathrm{d}\omega^2}\right|_{\omega_{0}} N(z)\right)\\
f_2(z)&=&\left(t-t_0- \frac{z}{c} + \frac{\pi}{
\lambda}\left.\frac{\mathrm{d}\chi_1}{\mathrm{d}\omega}\right|_{\omega_{0}}N(z)\right)\\
\bar{\alpha}(z)&=&\left( -\frac{i \pi}{\lambda}
\chi_1(\omega_{0})N(z)\right).
\end{eqnarray}
This completes our analytic exact solution. Though it contains an
infinite series, not all terms would be of significance at
condensate temperatures of interest. Furthermore one can still
make the result of more practical value by noting that it contains
a special function. In addition to its series and asymptotic
expansions, there are elementary, Gaussian-like functions that can
be fit to the error function. Thus these facts encourage us to
look for a semi-analytical method in which we fit polynomials or
Gaussian to the $N(z)$. Equivalently and more simply then one can
make such fits to the density profile of the condensate. Its exact
form is not essential as it only appears as an integrand. Due to
the approximate fitting involved, this method would be a
semi-analytical method to determine the pulse envelope. After
quickly reviewing standard numerical solvers of wave equation such
as Cranck-Nicolson and spectral methods, we shall test the
semi-analytical method against them.
\section{Typical Numerical methods for pulse propagation}
\subsection{Crank-Nicolson Method}
A dimensionless form of Eq. (\ref{eq:pulse}) can be solved via
finite difference Crank-Nicolson space marching scheme. The
Crank-Nicolson scheme is less stable but more accurate than the
fully implicit method; it takes the average between the implicit
and the explicit schemes \cite{Garcia}. Discritization is
performed as follows, with $i,j$ being the space and time grid
variables $(i,j=0,1,...,N)$ and  $E(z,t)\equiv E_j^i$,
\begin{eqnarray}
\frac{\partial E}{\partial z} &=& \frac{E_j^{i+1} -
E_j^{i}}{\Delta z};\quad\quad \frac{\partial E}{\partial t}=
\frac{E_{j+1}^{i} - E_{j-1}^{i}}{2 \Delta t}; \nonumber \\
\frac{\partial^2 E}{\partial t^2} &=& \frac{1}{2 (\Delta t)^2}(
[E_{j+1}^i + E_{j-1}^i - 2E_j^i] \nonumber\\&+& [E_{j+1}^{i+1} +
E_{j-1}^{i+1} - 2E_j^{i+1}] ). \label{eq:pulsewidth}
\end{eqnarray}
If we plug them into the wave equation (\ref{eq:pulse}), we get
\begin{eqnarray}
\frac{ib_2}{2  (\Delta t)^2} E_{j-1}^{i+1} + \left(\frac{1}{\Delta
z}- \frac{ib_2}{
 (\Delta t)^2}\right) E_{j}^{i+1} + \frac{ib_2}{2  (\Delta
t)^2}E_{j+1}^{i+1} \nonumber\\
 =
\left(-\frac{1}{2 v_g \Delta t} - \frac{ib_2}{2 (\Delta
t)^2}\right) E_{j-1}^{i} + \nonumber\\ \left(\alpha -
\frac{1}{\Delta z} - \frac{ib_2}{ (\Delta t)^2z}\right)
 E_{j}^{i} + \left(\frac{ib_2}{2 (\Delta t)^2}+\frac{1}{2v_g \Delta t}\right)
 E_{j+1}^{i} \label{matrixform}
\end{eqnarray}
By adding the boundary conditions (in our work set as zero:
$E_{0}^{i} = E_{N-1}^{i}=0$ for all $"i"$) we obtain set of $N$
linear equations with $N$ unknowns, which have to be solved
simultaneously for every space step $i$ where the vector $\{E_0\}$
is defined by initial conditions. Discrete equations in matrix
form are solved using Thomas algorithm \cite{Garcia} which is a
fast Gaussian elimination method for tridiagonal matrices.
\subsection{Pseudo-spectral Method}

Instead of doing a finite difference approximation in time, we can
expand the function $E(z,t)$ in spectral series at a given
position for all time values for better approximation of the time
derivative. The initial value can be used to determine the
coefficients of spectral series \cite{Boyd,Tref}. An appropriate
spectral series can be Fourier series. The reason we choose
Fourier series instead of polynomials or any other series that the
derivative of the Fourier series is just multiplication of Fourier
series with a pre-defined vector and also fast Fourier algorithm
makes it faster to compute the Fourier coefficients.

The initial function is divided into N points ($E_j$) and discrete
Fourier transform is applied in the interval $-N/2\leq k\leq
N/2-1$.
\begin{equation}
E_k=\mathcal{F} \{E_j\}=\frac{1}{N}\sum_{j=0}^{N-1}E_j e^{-i2\pi
kj/N} \quad
\end{equation}
for all k. We get the following ordinary differential equation;
\begin{equation}
\frac{\partial E(z,k)}{\partial
z}=-\left(\alpha+\frac{ik}{v_g}-ib_2 k^2 \right)E(z,k)
\end{equation}
We solve this ordinary differential equation for discrete space
intervals assuming that for each interval coefficients are
constants;
\begin{equation}
E(z_0+\Delta z,k)=E(z=z_0,k)\cdot e^{\Delta z(-\alpha -
\frac{ik}{v_g} + ib_2 k^2)}
\end{equation}
\section{Results and discussion}
\label{sec:results}

In our numerical calculations, we specifically consider a gas of
$N=8.3\times10^6$ $^{23}$Na atoms for which $M=23$ amu, $\lambda_0
= 589$ nm, $\gamma = 2\pi\times 10.01$ MHz, $\Gamma_3=0.5\gamma$,
$\Gamma_2=2\pi\times 10^3$ Hz, and $a_{s}=2.75$ nm. For the
parameters of the trapping potential, we take
$\omega_{r}=2\pi\times69$ Hz and $\omega_{z}=2\pi\times21$ Hz as
in Ref. \cite{slowlight-exp1}. The coupling field Rabi frequency
is taken to be $\Omega_c=0.56\gamma$ \cite{slowlight-exp1}.
Critical temperature for Bose-Einstein condensation of such a gas
is found to be $T_C=424$ nK.

\begin{figure}[htbp]
\centering{\vspace{0.5cm}}
\includegraphics[width=8.5cm]{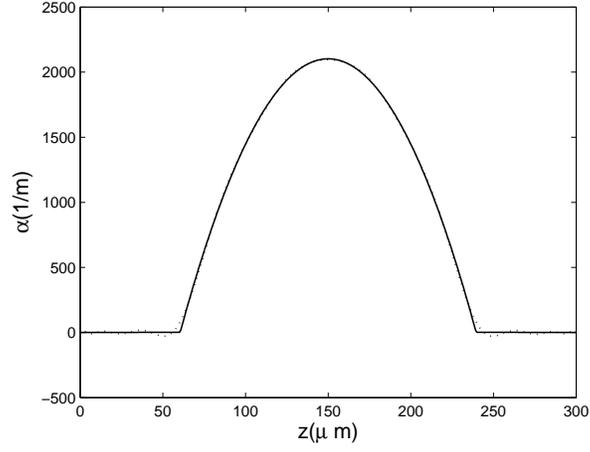}
\caption{Solid line shows the position dependence of the
absorption coefficient $\alpha$ along the $z$-axis. The dot line
is polynomial fitting for the absorption coefficient and the
degree of the polynomial is 22.  The ultracold atomic system of
$^{23}$Na $N=8.3\times10^6$ atoms at $T=42$ nK under EIT scheme.
The parameters used are $M=23$ amu, $a_{s}=2.75$ nm, $\lambda_0 =
589$ nm, $\gamma = 2\pi\times 10.01$ MHz, $\Gamma_3=0.5\gamma$,
$\Omega_c=0.56\gamma$, $\Gamma_2=2\pi\times 10^3$ Hz. }
\label{fig1}
\end{figure}

\begin{figure}[htbp]
\centering{\vspace{0.5cm}}
\includegraphics[width=3.25in]{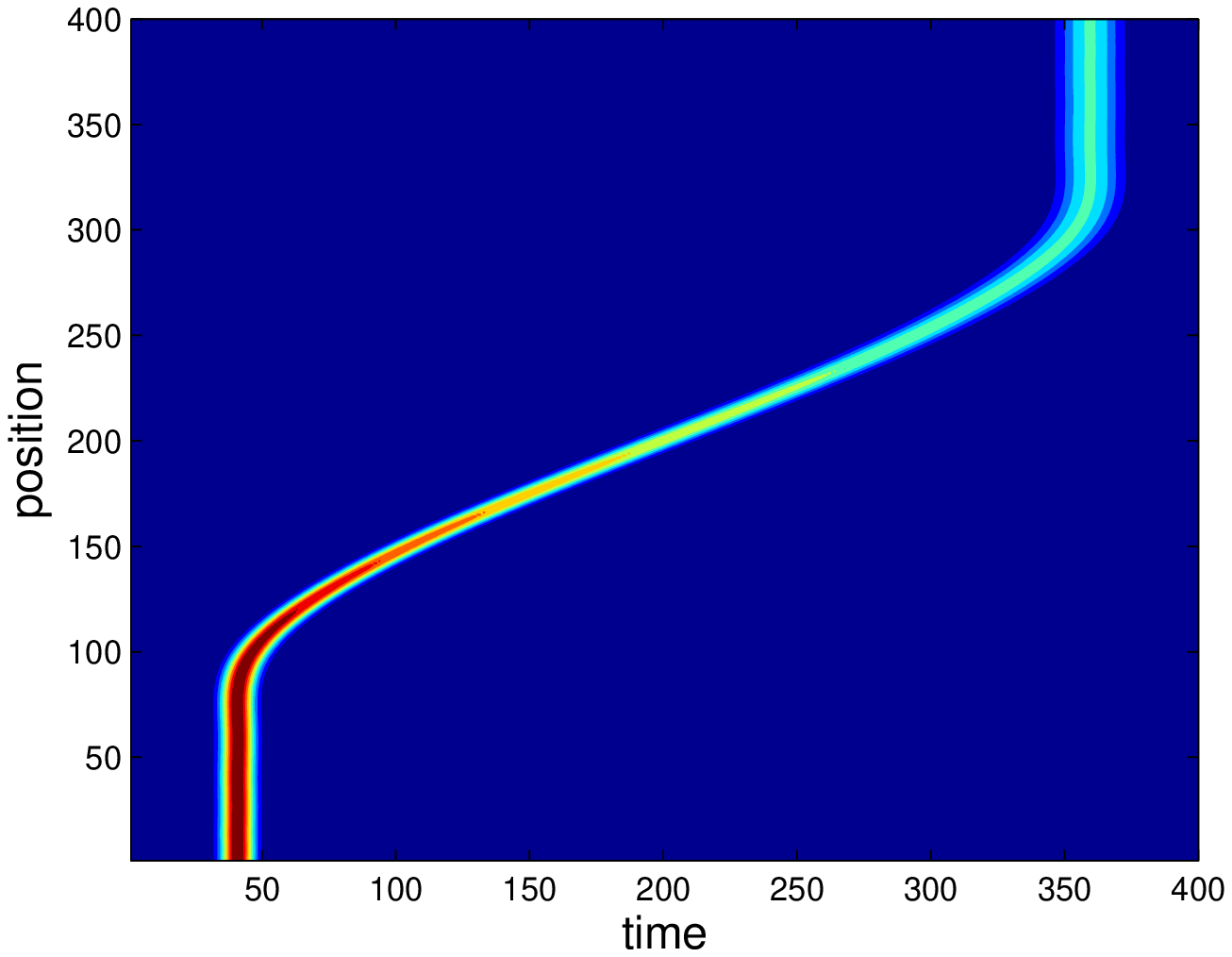}
\caption{Contour graph of the propagation of a microsecond pulse
through the interacting BEC by semi-analytical method. Time ($t$)
is scaled by $0.22\,\mu$s. and position ($z$) is scaled by
$1\,\mu$m. The parameters are the same with those of
Fig.\ref{fig1}. } \label{fig2}
\end{figure}

\begin{figure}[htbp]
\centering{\vspace{0.5cm}}
\includegraphics[width=3.25in]{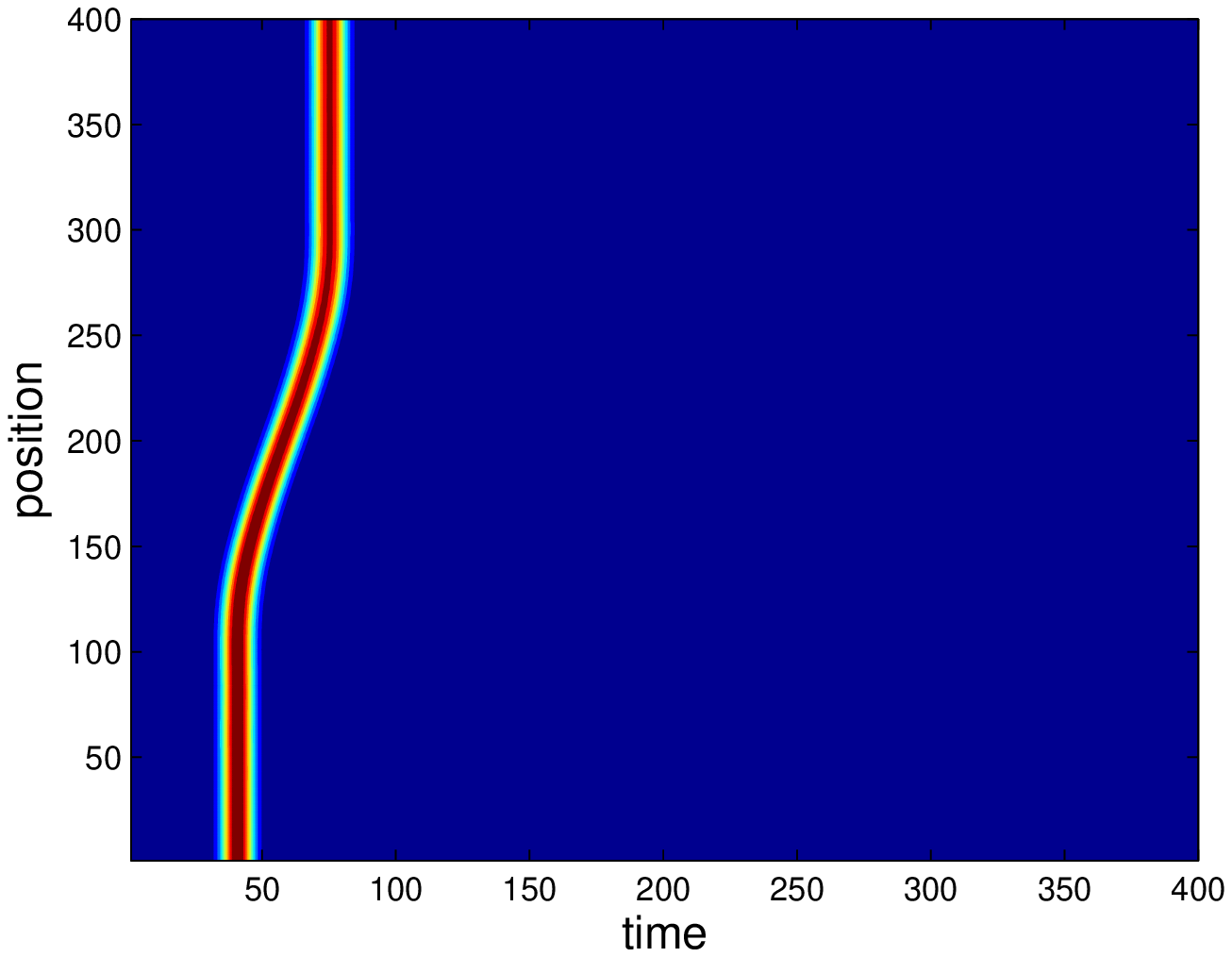}
\caption{Contour graph of the propagation of a microsecond pulse
through the interacting BEC by semi-analytical method. Time ($t$)
is scaled by $0.22 \,\mu$sec. and position ($z$) is by $1\,\mu$m.
The parameters are the same with those of Fig.\ref{fig1} except
for $\Omega_c=1.5\gamma$. } \label{fig3}
\end{figure}

Fitting a polynomial of degree $22$, pulse envelope is calculated
using the analytical formulae. To illustrate the success of the
fit we present the absorption coefficient in Fig.(\ref{fig1}) at
temperature $T=42$ nK. Solid line is the exact analytical
absorption coefficient while the dot line is the semi-analytical
result obtained after the polynomial fit. Similar behavior occurs
for $b_2$ and $1/v_g$. With the polynomial expressions of the
optical parameters, the integrals required for pulse envelope
functions such as $f_{1,2}$, or $N(z)$, are evaluated quickly and
simply. Contour plots of the propagating pulse are shown in Figs.
\ref{fig2}-\ref{fig3}. We assume a Gaussian pulse with unit
amplitude of the form $\exp{(-(t-t_0)^2 / 2 \tau_0^2)}$ at initial
time $t_0$, , where $\tau_0$ is the pulse width.
%
%
When the optical pulse enters the condensate region, its group
speed dramatically reduces under EIT conditions, as shown in Fig.
\ref{fig2}. The optical pulse rapidly assumes its high speeds
again following the passage to the thermal background cloud. We
assumed the optical pulse is propagating in vacuum before and
after the thermal component of the ultracold atomic system. In
experiment, group velocity is measured in terms of time delay of
the pulse with respect to a reference pulse which propagates in
vacuum over the same distance with the atomic medium. Broadening
of the pulse after leaving the condensate is visibly seen in the
figure as it gets about almost two times broader.

%
%

%
%

Similar behavior of the pulse but with a significant difference
regarding the pulse width can be seen in the Fig. \ref{fig3}. The
only change in the parameters used in Fig. \ref{fig2} is that now
$\Omega_c=1.5 \gamma$. In that case the broadening and absorption
becomes negligible while the pulse gets faster. These results, in
particular the role of the intensity of the control field on the
dispersion and loss management have already been discussed before
in Ref. \cite{tarhan2}. Here, the semi-analytical method is shown
to reproduce them efficiently.

To test the semi-analytical method with polynomial fitting against
standard numerical solvers of the wave equation, a dimensionless
form of Eq. (\ref{eq:pulse}) is also solved via finite difference
Crank-Nicolson (C-N) space marching scheme and pseudo-spectral
method (P-S). Final pulse width and final amplitude determined by
these methods are listed in Tab.(\ref{tab1}). A microsecond pulse
broadens by a factor of approximately $\sim 1.7$ according both
numerical and semi-analytical methods as seen in Tab.(\ref{tab1}).
Similar agreement of the semi-analytical method with the numerical
results are found for the absorption loss.

As a further test on our numerical methods as well, we have
compared the results of the Crank-Nicolson and pseudo-spectral
codes against the exact analytical solution for a uniform density
condensate cloud of $\rho=1.56 \times 10^{20}$ 1/m$^3$. The
coefficients in the Eq. (\ref{eq:pulse}) in that case are found to
be $v_g=1.5$ m/s, $\alpha=2.1\times 10^3$ 1/m, and $b_2=3.39
\times 10^{-8}$ s$^2$/m. Initial value for pulse width is $1\,
\mu$sec. The final pulse width determined according to the
Crank-Nicolson, pseudo-spectral, and completely analytic methods
are $3.3863\,\mu$sec, $3.3861\,\mu$sec, and $3.3869\,\mu$sec
respectively.

We can also use Gaussian fitting functions instead of polynomials
in order to get more explicit and compact expressions. Fitting a
Gaussian to the density for the same experimental parameters, we
have found the optical parameters become
\begin{eqnarray}
\alpha(z)=2.25\times10^{-3}\exp{\left[{-\frac{(z-1.5\times10^{-4})^2}{0.25\times10^{-8}}}\right]},
\nonumber \\ b_2(z)=-1.15\times10^{-8}\exp{\left[{-\frac{(z-1.5\times10^{-4})^2}{0.25\times10^{-8}}}\right]}, \\
1/v_g(z)=0.71\times
\exp{\left[{-\frac{(z-1.5\times10^{-4})^2}{0.25\times10^{-8}}}\right]}.
\nonumber
\end{eqnarray}
Reliability of the Gaussian fitting in the semi-analytical method
is tested against spectral method and Crank-Nicolson method as
summarized in Tab.(\ref{tab2}) and Tab.(\ref{tab3}). Results
obtained with numerical or semi-analytical Gaussian fit methods
are found to be in good agreement among themselves. Furthermore,
Gaussian fit method gives the similar results obtained with the
polynomial fit. This should be the case as the density only enters
as an integrand and the exact shape of the density should not be
essential to determine the optical pulse properties. Gaussian fit
method gives the final amplitude and width to be $0.475$ and
$1.7524\,\mu$sec, respectively. In these tables we have seen that
the agreement between the semi-analytical method and the numerical
methods seems to improve as the grid made finer and finer. This
may suggest that the semi-analytical method is more accurate than
the standard numerical solvers. However we are unable to give more
rigorous proof for that statement than these numerical tests.

Having analytical pulse envelope expressions or its reliable and
compact semi-analytical form allows us to vary controllable
experimental parameters easily to optimize pulse width, group
velocity and absorption loss of the pulse. To illustrate such
optimum dispersion management, we choose to examine the effect of
the trap curvature to illustrate. Effect of the control field
intensity has been numerically investigated earlier
\cite{tarhan2}. Considering a quadratic trap as $V=\kappa z^2$,
trap curvature is translated to the curvature of the dielectric
function through the density dependent EIT susceptibility. Typical
results obtained by the Gaussian fitting semi-analytical method
are listed in listed in Tab.(\ref{tab4}). The parameters are the
same with those of Fig.\ref{fig1} and $\kappa$ is in units of
kg/sec$^2$. It is seen that pulse shape is better preserved at
larger $\kappa$.

\section{Conclusion}
\label{sec:conclusion}

We have discussed analytical and semi-analytical solution of the
one -dimensional wave equation which governs the propagation of an
ultraslow optical pulse in a dispersive inhomogeneous  atomic
condensate. Ignoring paraxial effects, slowly varying pulse
envelope wave equation is solved by using Fourier transformation
technique and by using special functions, in particular, normal
cumulative distribution function which is related to the error
function. It has been argued that, as the exact density profile
only enters as an integrand to the expressions, simpler polynomial
and Gaussian fits can be made for more compact and simpler
expressions. Such a semi-analytical method is heavily tested
against standard numerical solvers of the wave equation, namely,
Cranck-Nicolson and pseudo-spectral methods. The results confirm
the reliability of the compact expressions obtained under
semi-analytical method. As an illustration of the efficiency of
analytical expressions, the role of trap curvature on the
dispersion management has been investigated. A quick calculation
shows that pulse shape is more preserved in traps with higher
curvatures. Other experimentally controllable parameters can be
similarly studied for optimized design of optical traps and EIT
conditions to efficient storage of coherent optical information or
to engineer ultraslow pulse shapes.

%

\newpage

\section*{List of Figure Captions}

\textbf{Fig. 1.} Solid line shows the position dependence of the
absorption coefficient $\alpha$ along the $z$-axis. The dot line
is polynomial fitting for the absorption coefficient and the
degree of the polynomial is 22.  The ultracold atomic system of
$^{23}$Na $N=8.3\times10^6$ atoms at $T=42$ nK under EIT scheme.
The parameters used are $M=23$ amu, $a_{s}=2.75$ nm, $\lambda_0 =
589$ nm, $\gamma = 2\pi\times 10.01$ MHz, $\Gamma_3=0.5\gamma$,
$\Omega_c=0.56\gamma$, $\Gamma_2=2\pi\times 10^3$ Hz.

\textbf{Fig. 2.} Contour graph of the propagation of a microsecond
pulse through the interacting BEC by semi-analytical method. Time
($t$) is scaled by $0.22\,\mu$s. and position ($z$) is scaled by
$1\,\mu$m. The parameters are the same with those of
Fig.\ref{fig1}

\textbf{Fig. 3.} Contour graph of the propagation of a microsecond
pulse through the interacting BEC by semi-analytical method. Time
($t$) is scaled by $0.22 \,\mu$sec. and position ($z$) is by
$1\,\mu$m. The parameters are the same with those of
Fig.\ref{fig1} except for $\Omega_c=1.5\gamma$.

\newpage
\section*{List of Table}
\begin{table}[ht]
\centering \caption{Comparison of Crank-Nicolson (C-N),
pseudo-spectral (P-S), and semi-analytical (S-A) method.
Propagation of optical pulse with initial pulse width $1 \,\mu$sec
and initial amplitude $1$.}
{\begin{tabular}{@{}cccc@{}} \hline  & C-N & P-S & S-A   \\
\hline \hline
final pulse width & $1.7311\, \mu$sec\hphantom{0} & $1.7305\, \mu$sec\hphantom{0}  & $1.7309\, \mu$sec\hphantom{0} \\
final amplitude &0.4755\hphantom{0} & 0.4754\hphantom{0} & 0.4758\hphantom{0} \\
\hline
\end{tabular} } \label{tab1}
\end{table}
\begin{table}[ht]
\centering \caption{Ratio of the results for pseudo-spectral
method to the semi-analytical values for the given
position$\times$time grid dimensions.} {\begin{tabular}{@{}ccc@{}}
\hline grid dimensions &\,\,final amplitude &\,\,final width  \\
\hline \hline
 $2^9\times2^9$ & 0.998 & 0.9993   \\
 $2^{10}\times2^{10}$ & 0.9993 & 0.9997   \\
 $2^{12}\times2^{10}$ & 0.9995 & 0.9998   \\
 $2^{12}\times2^{12}$ & 0.9998 & 0.99991  \\
\hline
\end{tabular} }  \label{tab2}
\end{table}
\begin{table}[ht]
\centering \caption{Ratio of the results for Crank-Nicolson method
to the semi-analytical values for the given position$\times$time
grid dimensions.}
{\begin{tabular}{@{}ccc@{}}  \hline grid dimensions &\,\,final amplitude &\,\,final width  \\
\hline \hline
 $2^9\times 2^9 $ & 0.9998 & 0.9994   \\
 $2^{12}\times2^{10}$ & 0.99988 & 0.99991   \\
 $2^{12}\times2^{12}$ & 0.99998 & 0.99994  \\
\hline
\end{tabular} }  \label{tab3}
\end{table}

\newpage
\begin{table}[ht]
\centering \caption{Dispersive propagation of optical pulse as a
function of $\kappa$.}
{\begin{tabular}{@{}ccc@{}} \hline  $\kappa$ &final amplitude &final width ($\mu$sec)\\
\hline \hline
 $2\times10^{-22}$ & 0.4113 &  1.8916  \\
 $0.5\times10^{-21}$ & 0.5250 &  1.6247 \\
$1\times10^{-21}$ & 0.6121 & 1.4612  \\
 $0.5\times10^{-20}$ & 0.7629 & 1.2422  \\
 $1\times10^{-20}$ &0.8212  & 1.1741   \\
 \hline
\end{tabular} }  \label{tab4}
\end{table}


\newpage
\newpage
\newpage
\end{document}